\documentstyle[12pt]{article}
\input epsf

\def\lsim{\mathrel{\rlap{\lower 4pt \hbox{\hskip 1pt $\sim$}}\raise 
1pt\hbox{$<$}}}
\def\gsim{\mathrel{\rlap{\lower 4pt \hbox{\hskip 1pt $\sim$}}\raise 
1pt\hbox{$>$}}}

\begin{document}
\begin{titlepage}
\begin{center}
\vspace{-1.5in}
\LARGE

\def\thefootnote{\fnsymbol{footnote}}

{\bf Observational tests of inflation}\footnote{To appear, 
	proceedings of {\em Inner Space, Outer Space II}, Fermilab, May 1999.}\\
\vspace{.3in}
\normalsize
\large{Andrew R.~Liddle} \\
\normalsize
\vspace{.6 cm}
{\em Astrophysics Group,\\ The Blackett Laboratory,\\ Imperial College,\\
 London SW7 2BZ, UK}

\end{center}

\setcounter{footnote}{0}

\vspace{.6 cm}

\section*{Abstract} 
\noindent  
We are on the verge of the first precision testing of the inflationary
cosmology as a model for the origin of structure in the Universe. I
review the key predictions of inflation which can be used as
observational tests, in the sense of allowing inflation to be falsified.
The most important prediction of this type is that the perturbations
will cross inside the Hubble radius entirely in their growing mode,
though nongaussianity can also provide critical tests.  Spatial flatness and
tensor perturbations may offer strong support to inflation, but cannot be used
to exclude it. Finally, I discuss the extent to which observations will
distinguish between inflation models, should the paradigm survive these
key tests, in particular describing a technique for reconstruction of the 
inflaton potential which does not require the slow-roll approximation.

%%%%%%%%%%%%%%%%%%%%%%%%%%%%%%%%%%%%%%%%%%%%%%%%%%%%%%%%%%%%%%%%%%%%%
\end{titlepage}
%%%%%%%%%%%%%%%%%%%%%%%%%%%%%%%%%%%%%%%%%%%%%%%%%%%%%%%%%%%%%%%%%%%%%%
%\double %Removal of the first `%' generates a double spaced copy

\section{What does inflation predict?}

Cosmological inflation \cite{inf,LL,LR} is widely perceived as an excellent 
paradigm
within which one can explain both the global properties of the
Universe and the irregularities which give rise to structures within
it. Despite this, it remains fair to say that as yet the inflationary
paradigm has been confronted with only a few observational challenges,
which it has comfortably surmounted. In years to come, it will face many
more, and the purpose of this article is to discuss which of these tests
are likely to be the most stringent.

In doing so, it is worthwhile to separate out the two key roles that
inflation plays in modern cosmology. The first, which led to its
introduction, is in setting the `initial conditions' for the global
Universe, by arranging a large homogeneous Universe devoid of unwanted
relics such as monopoles. In terms of these global properties, it now
seems unlikely that any new observations will undermine the inflationary
picture, and, as Linde has argued \cite{Lincrit}, if it is to be supplanted that 
is
likely to be because of the advent of a superior theory, rather than of
superior observations. Accordingly, I will have little to say on this
topic.

The second role, which is potentially much more fruitful as a probe of
high-energy physics, is that inflation provides a theory for the origin
of perturbations in the Universe (for reviews, see Refs.~\cite{LL,LLKCBA}). As 
these perturbations are believed to
evolve into all the observed structures in the present Universe, including
the existence and clustering of galaxies and the anisotropies in the
cosmic background radiation, this proposal is subject to a wide variety of
observational tests. Thus far, these tests have been rather qualitative
in nature, but in the near future inflation as a theory of the origin
of structure in the Universe will face precision testing.

The challenge facing cosmologists is therefore to address two questions:
\begin{itemize}
\item Is inflation right?
\item If so, which version of inflation is right?
\end{itemize}
Unfortunately, in science one never gets to prove that a theory is
correct, merely that it is the best available explanation. The way
to convince the community that a theory is indeed the best explanation
is if that theory can repeatedly pass new observational tests. In that
regard, it is important to be as clear as possible concerning what these
tests might be.

\section{The predictions of inflation}

The essence of testing inflation can be condensed into a single
sentence, namely
\begin{quote}
The {\em simplest} models of inflation predict
\underline{power-law} spectra of \underline{gaussian}
\underline{adiabatic} scalar and \underline{tensor} perturbations in
their \underline{growing mode} in a \underline{spatially-flat}
Universe.
\end{quote}
This sentence contains 6 key predictions of the inflationary paradigm,
which I've underlined, but also one crucial word, `simplest'. The
trouble is that inflation is a paradigm rather than a model, and has
many different realizations which can lead to a range of different
predictions. From a straw poll of cosmologists, everyone agreed that
there were at least several tens of different models, and I'd say there
certainly aren't as many as a thousand, so a reasonable first guess is
that at present there are around one hundred different models on the
market, all consistent (at least more or less) with present
observational data.

A valuable scientific theory is one which has sufficient predictive
power that it can be subjected to observational tests which are
capable of falsifying it. When the model survives such a test, it
strengthens our view that the model is correct; in Bayesian terms, its
likelihood is increased relative to models which are less capable of
matching the data. It is useful to think of these models at three
different levels: 
\begin{itemize}
\item {\bf Specific models of inflation.} These are readily testable. For 
example, there will be a specific prediction for the spectral index
$n$ of the density perturbations, and this will be measured to high
accuracy. 
\item {\bf Classes of models.} This means models sharing some common
property, for example that the perturbations are gaussian. An entire
class of models can be excluded by evidence against that shared
property.
\item {\bf The inflationary paradigm.} Testing the paradigm as a whole
requires a property which is robust amongst all models. As we'll see,
the most striking such property is that the perturbations should be in
their growing mode, which leads to the distinctive signature of
oscillations in the microwave anisotropy power spectra.
\end{itemize}
Finally, note that since one can never completely rule out a small
inflationary component added on to some rival structure formation model (e.g.~a 
combined cosmic strings and inflation model \cite{chm}),
in practice we are initially testing the paradigm of inflation as the {\em
sole} origin of structure in the Universe.

It can also be helpful to make the admittedly rather narrow distinction
between tests and supporting evidence \cite{Peebles}. A test arises when
there is a prediction which, if contradicted by observations, rules out
the model, or at least greatly reduces its likelihood relative to a
rival model. In this sense, the geometry of the Universe is not a test
of inflation, because there exist inflation models predicting whatever
geometry might be measured (including open and closed ones), and there is no 
rival regarded as giving a better explanation for any particular possible 
observation. By contrast, the
oscillations in the microwave anisotropy power spectra (both temperature and 
polarization) do give rise to a
test, as we will shortly see. 

Supporting evidence arises with observational confirmation of a prediction which
is regarded as characteristic, but which is not generic.  A good example would
be the observation of tensor perturbations with wavelengths exceeding the Hubble
length, for which inflation would be by far the best available explanation; they
do not give rise to a test because if they are not observed, then there are
plentiful inflation models where such perturbations are predicted to be below
any anticipated observational sensitivity.

\section{Testing the predictions}

\subsection{Spatial flatness}

Of the listed properties, spatial flatness is the only one which
refers to the global properties of the Universe.\footnote{Inflation is
also responsible for solving the horizon problem, ensuring a Universe
close to homogeneity, but this is no longer a useful test as it is
already observationally verified to high accuracy through the
near isotropy of the cosmic microwave background.} It is particularly
pertinent because of the original strong statements that spatial
flatness was an inevitable prediction of inflation, later retracted
with the discovery of a class of models --- the open inflation models
\cite{open,HT}
--- which cunningly utilize quantum tunnelling to generate homogeneous
open Universes. In the recent `tunnelling from nothing' instanton models
of Hawking and Turok \cite{HT}, any observed curvature has the interesting
interpretation of being a relic from the initial formation of the
Universe which managed to survive the inflationary epoch.

If we convince ourselves that, to a high degree of accuracy, the
Universe is spatially flat, that will strengthen the likelihood that the
simplest models of inflation are correct. However, an accurate
measurement of the curvature is not a {\em test} of the full
inflationary paradigm, because whatever the outcome of such a
measurement there do remain inflation models which make that prediction.
This point has recently been stressed by Peebles \cite{Peebles}. The likelihood 
will
have shifted to favour some inflation models at the expense of others,
but the total likelihood of inflation will be
unchanged.\footnote{Indeed, the only existing alternative to inflation
in explaining spatial flatness is the variable-speed-of-light (VSL)
theories \cite{vsl}, which may be able to solve the problem
without inflation, though at the cost of abandoning Lorentz invariance.
There are no available alternatives at all to inflation in explaining an
open Universe, so one might say that observation of negative curvature
modestly {\em improves} the likelihood of inflation amongst known
theories, by eliminating the VSL theories from consideration.} Only if a
rival class of theories can be invented, which predict say a
negative-curvature Universe in a way regarded as more compelling than
the open inflation models, will measurements of the curvature acquire
the power to test the inflationary paradigm.

I should also mention that the standard definition of inflation --- a
period where the scale factor $a(t)$ undergoes accelerated expansion
--- is a rather general one, and in particular any classical solution
to the flatness problem using general relativity must involve
inflation. This follows directly from writing the Friedmann equation
as
\begin{equation}
|\Omega - 1| = \frac{|k|}{\dot{a}^2} \,.
\end{equation} 
An example is the pre big bang cosmology \cite{PBB}, which is now viewed as a
novel type of inflation model rather than a separate idea. This makes
it hard to devise alternative solutions to the flatness problem; open
inflation models use quantum tunnelling but in fact still require
classical inflation after the tunnelling, and presently the only
existing alternative is the variable-speed-of-light theories \cite{vsl} which 
violate general relativity. 

Before continuing on to the properties of perturbations in the Universe,
there's a final point worth bearing in mind concerning inflation as a
theory of the global Universe. As I've said, there now seems little
prospect that any observations will come along which might rule out the
model. But it is interesting that while that is true now, it was {\em
not} true when inflation was first devised. An example is the question
of the topology of the Universe.  We now know that if there is any
non-trivial topology to the Universe, the identification scale is at
least of order the Hubble radius, and I expect that that can be
consistent with inflation (though I am unaware of any detailed
investigation of the issue). However, from observations available in
1981 it was perfectly possible that the identification scale could have
been much much smaller. Since inflation will stretch the topological
identification scale, that would have set an upper limit on the amount
of inflation strong enough to prevent it from solving the horizon and
flatness problems. The prediction of no small-scale topological
identification has proven a successful one. Another example of a test
that could have excluded inflation, but didn't, is the now-observed
absence of a global rotation of our observable Universe \cite{BL}.

\subsection{Growing-mode perturbations}

This is the key prediction of inflation as a theory of the origin of
structure; inflation generically predicts oscillations in the
temperature and polarization angular power spectra. If oscillations are not
seen, then inflation cannot be the sole origin of structure.

The reason this prediction is so generic is because inflation creates
the perturbations during the early history of the Universe, and they
then evolve passively until they enter the horizon in the recent past.
The perturbations obey second-order differential equations which possess
growing and decaying mode solutions, and by the time the perturbations
enter the horizon the growing mode has become completely dominant. That
means the solution depends only on one parameter, the amplitude of the
growing mode; in particular, the derivative of the perturbation is a
known function of the amplitude. The solution inside the horizon is
oscillatory before decoupling, and this fixes the temporal phase of the
perturbations; all perturbations of a given wavenumber oscillate
together and in particular at any given time there are scales on which
the perturbation vanishes. Projected onto the microwave sky, this leads to the 
familiar peak structure
seen in predicted anisotropy spectra, though if one wants to be pedantic the 
troughs are if anything more significant than the peaks. 

The importance of the peak structure in distinguishing inflation from
rivals such as defect theories was stressed by Albrecht and
collaborators \cite{Aetal} and by Hu \& White \cite{HW}. The prediction
is a powerful one; in particular it still holds if the inflationary
perturbations are partly or completely isocurvature, and if they are
nongaussian. 

I stress that while inflation inevitably leads to the oscillations, I am
not saying that inflation is the {\em only} way to obtain oscillations. For
example there are known active source models which give an oscillatory
structure \cite{Neil}, though the favourite cosmic string model is believed not 
to.
If observed, oscillations would support inflation but cannot prove it.
However I might mention in passing that it is quite easy to prove
\cite{L95} that the existence of adiabatic perturbations on scales much
larger than the Hubble radius would imply one of three possibilities;
the perturbations were there as initial conditions, causality/Lorentz
invariance is violated, or a period of inflation occurred in the past.

\subsubsection*{Against designer inflation}

At this point it is worth saying something against `designer' models of
inflation which aim to match observations through the insertion of
features in the spectra, by putting features in the
inflationary potential. This idea first arose in considering the matter
power spectrum \cite{SBB}, which is a featureless curve and so quite amenable to
the insertion of peaks and troughs. However the idea is much more
problematic in the context of the microwave anisotropy spectra, which
themselves contain sharp oscillatory features. One might 
contemplate inserting oscillations into the initial power spectrum in
such a way as to `cancel out' the oscillatory structure, but there are
however three levels of argument against this:
\begin{enumerate}
\item It's a silly idea, because the physics during inflation has no
idea where the peaks might appear at decoupling, and for the idea to be
useful they have to match to very high accuracy. That argument is good
enough for me, though perhaps not for everyone, so ...
\item Even if you wanted to do it you probably cannot. Barrow and myself
\cite{BL} found that the required oscillations were so sharp as to be
inconsistent with inflation taking place, at least in single-field inflation
models. However, a watertight case remains to be made on this point, and
it is not clear how one could extend that claim to multi-field models,
so perhaps the most pertinent argument is ...
\item Even if you managed to cancel out the oscillations in the
temperature power spectrum, the polarization spectra have oscillations
which are out of phase with the temperature spectrum, and so those
oscillations will be enhanced \cite{LA}.
\end{enumerate} 

\subsection{Gaussianity and adiabaticity}

While the simplest models of inflation predict gaussian adiabatic
perturbations, many models are known which violate either or both of
these conditions. Consequently there is no critical test of inflation
which can be simply stated. Nevertheless, it is clear that these could
lead to tests of the inflationary paradigm. For example, as far as
inflation is concerned, there is good nongaussianity and bad
nongaussianity. For example, if line discontinuities are seen in the
microwave background, it would be futile to try and explain them using
inflation rather than cosmic strings. On the other hand, nongaussianity
with a chi-squared distribution is very easy to generate in inflation
models; one only has to arrange that the leading contribution to the
density comes from the square of a scalar field
perturbation. Indeed, in isocurvature inflation models, it appears at
least as easy to arrange chi-squared statistics as it is to arrange
gaussian ones \cite{LM}.

Inflation may also be able to explain nongaussian perturbations of a
`bubbly' nature, by attributing the bubbles to a phase transition
bringing inflation to an end. The simplest models of this type have
already been excluded, but more complicated ones may still be viable.

\subsection{Tensor and vector perturbations}

Gravitational wave perturbations, also known as tensor perturbations, 
are inevitably produced at some level by inflation, but the level
depends on the model under consideration and it is perfectly possible,
and perhaps even likely \cite{Lyth97}, that the level is too small to be 
detected by
currently envisaged experiments. This prevents them acting as a test.

In standard inflation models, the gravitational waves are directly
observable only by the microwave background anisotropies they induce.
Assuming Einstein gravity, the Hubble parameter always decreases during
inflation which leads to a spectrum which decreases with decreasing
scale; the upper limit set by these anisotropies places the amplitude on
short scales orders of magnitude below planned detectors (and probably
well below the stochastic background from astrophysical sources). The
exception is the pre big bang class of models \cite{PBB} (implemented in 
extensions
of Einstein gravity), where the gravitational wave spectrum rises
sharply to short scales and is potentially visible in laser
interferometer experiments.

As well as the direct effect on the microwave background, gravitational
waves evidence themselves as a deficit of short-scale power in the
density perturbation spectrum of COBE-normalized models. Presently the
combination of large-scale structure data with COBE gives the strongest
upper limit on the fractional contribution $r$ on COBE scales, at $r
\lsim 0.5$ \cite{Zibin}. There is no evidence to favour the tensors, but
this constraint is fairly weak. Eventually the {\sc Planck} satellite is
expected to be able to detect (at 95\% confidence) a contribution above
$r \sim 0.1$ \cite{TEHC}, and may perhaps do better if there is early 
reionization and/or the foreground contamination turns out to be readily 
modellable.
Conceivably, high-precision observations of the polarization of the
microwave background might improve this further.

The verdict, therefore, is that if a tensor component is seen,
corresponding to gravitational waves on scales bigger than the Hubble
radius at decoupling, that is extremely powerful support for the
inflationary paradigm. This would be stronger yet if the observed
spectrum could be shown to satisfy an equation known as the consistency equation 
to some
reasonable accuracy; this relates the tensor spectral index to the 
relative amplitude of tensors and scalars, and signifies the common origin of 
the two spectra from a single inflationary potential $V(\phi)$ 
\cite{LL92,Recon}. However, the tensor perturbations do not provide
a test for inflation in the formal sense, since no damage is inflicted
upon the inflationary paradigm if they are not detected.

While known inflationary models generate both scalar and tensor modes,
it appears extremely hard to generate large-scale vector modes. There
are two obstacles. The first is that massless vector fields are
conformally invariant, which means that perturbations are not excited by
expansion; this has to be evaded either by introducing a mass (which
suppresses the effect of perturbations) or an explicit coupling breaking
conformal invariance \cite{vec}. The second obstacle is that vector 
perturbations die off
rapidly as the Universe expands, and to survive until horizon entry
their initial value would have to be considerably in excess of the
linear regime. In consequence, a significant prediction of inflation is
the absence of large-scale vector perturbations. If they are seen, it
seems likely to be impossible to make them with inflation alone, though
I am not aware of a cast-iron proof. By contrast, topological defect
models generically excite vector perturbations.

\section{Discriminating inflationary models}

\subsection{Power-law behaviour}

If the inflationary paradigm survives the tests above, it will be time
to decide which of the existing inflation models actually fits the data.
In most models, to a good approximation the density perturbations are
given by a power-law and the gravitational waves are at best marginally
detectable by {\sc Planck}. Accurate measures of these two quantities have the
potential to exclude nearly all existing inflation models.

At present, the spectral index is very loosely constrained; in general
the limits are probably around $0.8 < n < 1.3$, though if specific
assumptions are made (e.g.~critical matter density or significant
gravitational waves) this can tighten. As it happens, this entire viable range
is fairly well populated by inflation models, which means that any
increase in observational sensitivity has the power to exclude a
significant fraction of them.

A benchmark for future accuracy is the {\sc Planck} satellite; recently a
detailed analysis, including estimates of foreground removal efficiency,
concluded that it would reach a 1-sigma accuracy on $n$ of around $\pm 0.01$ 
\cite{TEHC}. By 
contrast,
the 1-sigma error from the MAP satellite is predicted to be in the range
$0.05$ to $0.1$, which in itself may not significantly impact on
inflationary models, though it may be powerful in combination with
probes such as the power spectrum from the Sloan Digital Sky Survey.

\subsection{Deviations from the power-law}

The power-law approximation to the spectra, as derived in
Ref.~\cite{LL92}, is particularly good at the moment because the
available observations are not very accurate. In most models the spectra are
indistinguishable from power laws even at {\sc Planck} accuracy, but there are
exceptions, and if deviations are observable they correspond to extra
available information on the inflationary spectrum \cite{KT,CGL}. One such class 
of
models are models where features have been deliberately inserted into
the potential in order to generate sharp features in the power spectrum,
such as the broken scale-invariance models. 

However, even without a specific feature, it may be possible to see
deviations, if the slow-roll approximation is not particularly good.
There is actually modest theoretical prejudice in favour of this,
because in supergravity models the inflaton is expected to receive
corrections to its mass which are large enough to threaten slow-roll \cite{LR}.
Specifically, the slow-roll parameter $\eta$, which is supposed to be
small, receives a contribution of
\begin{equation}
\eta = 1 + {\rm `something'} \,,
\end{equation}
where the `something' is model dependent. It is clear that if slow-roll
is to be very good, $\eta \ll 1$, then the `something' has to cancel the
`1' to quite high accuracy, and there is no theoretical motivation
saying it should.

If we accept that, then we conclude that $n$ should not be extremely
close to one (which would exacerbate the need for cancellation), and also
that the deviation from scale-invariance, $dn/d\ln k$, which is given by the 
slow-roll parameters, might be large enough to be
measurable \cite{CGL}. A specific example where this is indeed the case is the
running mass model \cite{Stew,CL}.

\subsection{Reconstruction without slow-roll}

Eventually, in order to get the best possible constraints on inflation
one will want to circumvent the slow-roll approximation completely, and
this can be done by computing the power spectra (first the scalar and
tensor spectra, and from them the induced microwave anisotropies) entirely 
numerically.
Such an approach was recently described by Grivell and Liddle
\cite{GL99}, and represents the optimal way to obtain constraints on
inflation from the data (though at present it has only been implemented
for single-field models). 

\begin{figure}
\centering
\leavevmode\epsfysize=9.3 cm \epsfbox{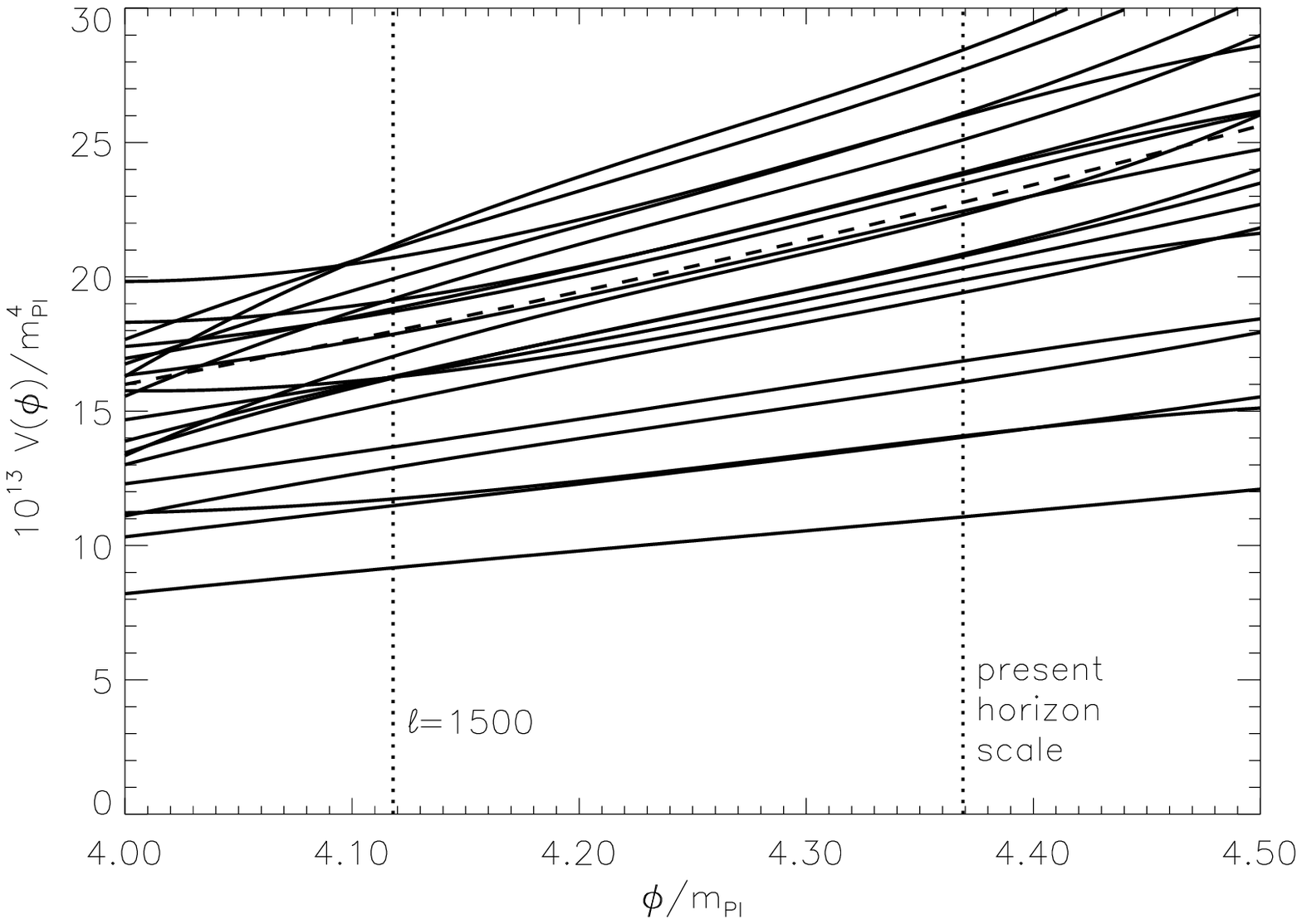}\\ 
\leavevmode\epsfysize=9.3 cm \epsfbox{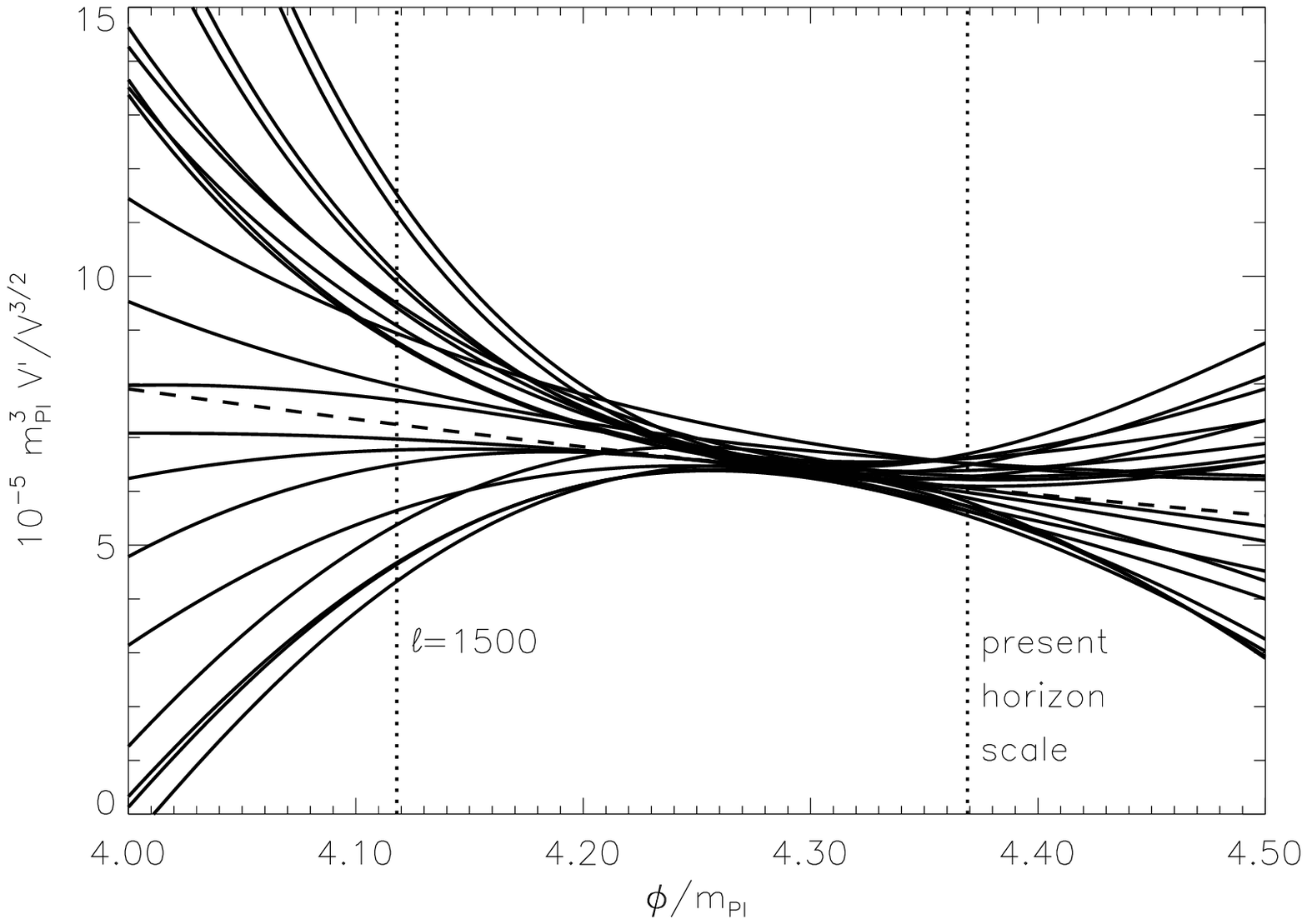}\\ 
\caption[recfig]{\label{f:recfig} The $\lambda \phi^4$ potential as seen by 
the {\sc Planck} satellite. In the upper panel, the dashed line shows the true 
potential, and the full lines show a series of Monte Carlo reconstructions, 
which differ in the realization of the observational errors. In reality we get 
only one of these curves. The lower panel shows the combination 
$V'/V^{3/2}$, which is much better determined. Scalar field values when scales 
equalled the Hubble radius during inflation are shown, roughly corresponding to 
the range of {\sc Planck}.}
\end{figure}

In this approach, rather than estimating quantities such as the spectral
index $n$ from the observations, one directly estimates the potential,
in some parametrization such as the coefficients of a Taylor series.
An example is shown in Figure~\ref{f:recfig}, which shows a test case of a
$\lambda \phi^4$ potential as it might be reconstructed by the {\sc Planck}
satellite --- see Ref.~\cite{GL99} for full details. Twenty different 
reconstructions are shown (corresponding to 
different realizations of the random observational errors), whereas in the real 
world we would get only one of these. We see considerable variation, which 
arises because the overall amplitude can only be fixed by detection of the 
tensor component, which is quite marginal in this model. However, there are 
functions of the potential which are quite well determined. The lower panel 
shows $V'/V^{3/2}$ (where the prime is a derivative with respect to the field), 
which is given to an accuracy of a few percent on the scales where the 
observations are most efficient. This particular combination is favoured because 
it is the combination which (at least in the slow-roll approximation) gives the 
density perturbation spectrum.

No doubt, when first confronted with quality data people will aim to determine 
$n$, $r$, and so on along with the cosmological parameters such as the density 
and Hubble parameter. However, if we become convinced that the inflationary 
explanation is a good one, this direct reconstruction approach takes maximum 
advantage of the data in constraining the inflaton potential.

\section{Outlook}

While the present situation is extremely rosy for inflation, which
stands as the favoured model for the origin of structure, there is a
sense in which the present is the worst time to be considering inflation
models. A quick survey of the literature suggests that there are perhaps
of order 100 viable models of inflation, the most there has ever been.
At the original {\em Inner Space, Outer Space} meeting in 1984, there were only
a handful. It's true that some models devised in those 15 years have been 
excluded, such as the extended inflation models \cite{LS,LL92}, but model 
builders have for the most part had quite a free hand operating within the given 
constraints.

Further, this is likely to be about the most viable models there will
ever be, because observations are at the threshold of significantly
impacting on this collection. Experiments such as Boomerang, VSA and MAP are
capable of ruling out inflation completely, by one of the methods
outlined in this article. If inflation survives, they will have
significantly reduced the number of models, and then a few years later
{\sc Planck} should eliminate most of the rest. Hopefully, by the time of
{\em Inner Space, Outer Space III} in 2014, we will be back once more to a
handful.

%%%%%%%%%%%%%%%%%%%%%%%%%%%%%%%%%%%%%%%%%%%%%%%%%%%%%%%%%%%%%%%%%%%%%%
\section*{Acknowledgments}

Thanks to Ed Copeland, Ian Grivell, Rocky Kolb and Jim Lidsey for
collaborations which underlie much of the discussion in this article,
and to Anthony Lasenby for stressing to me the importance of the absence of 
vector
perturbations as a prediction of known inflationary models. I thank the Royal 
Society for financial support.

%%%%%%%%%%%%%%%%%%%%%%%%%%%%%%%%%%%%%%%%%%%%%%%%%%%%%%%%%%%%%%%%%%%%%%

%%%%%%%%%%%%%%%%%%%%%%%%%%%%%%%%%%%%%%%%%%%%%%%%%%%%%%%%%%%%%%%%%%%%%%

\end{document}